\documentclass[journal,comsoc]{IEEEtran}
\usepackage{amsmath,amsfonts}
\usepackage{algorithmic}
\usepackage{algorithm}
\usepackage{array}
\usepackage[caption=false,font=normalsize,labelfont=sf,textfont=sf]{subfig}
\usepackage{textcomp}
\usepackage{stfloats}
\usepackage{url}
\usepackage{verbatim}
\usepackage{graphicx}
\usepackage{cite}
\usepackage{multirow}
\usepackage[table]{xcolor}
\usepackage{tikz}
\usepackage{eso-pic}

\newcommand\copyrighttext{%
  \footnotesize \textcopyright \the\year{} IEEE. Personal use of this material is permitted.  Permission from IEEE must be obtained for all other uses, in any current or future media, including reprinting/republishing this material for advertising or promotional purposes, creating new collective works, for resale or redistribution to servers or lists, or reuse of any copyrighted component of this work in other works. DOI: 10.1109/JIOT.2025.3550204}

\AddToShipoutPicture*{%
  \begin{tikzpicture}[remember picture,overlay]
    \node[anchor=south,yshift=10pt] at (current page.south) 
      {\fbox{\parbox{\dimexpr0.7\textwidth-\fboxsep-\fboxrule\relax}{\copyrighttext}}};
  \end{tikzpicture}%
}
\begin{document}

\title{Gait Disorder Assessment Based on\\a Large-Scale Clinical Trial:\\WiFi vs. Video vs. Doctor’s Visual Inspection}

\author{Alireza Parsay*,
\IEEEmembership{Member, IEEE}, Mert Torun*, \IEEEmembership{Member, IEEE}, Philip R. Delio, and Yasamin Mostofi,  \IEEEmembership{Fellow, IEEE}
\thanks{A. Parsay, M. Torun, and Y. Mostofi are with the Department of Electrical and Computer Engineering, University of California, Santa Barbara (e-mail: alirezaparsay@ucsb.edu; merttorun@ucsb.edu; ymostofi@ece.ucsb.edu).}
\thanks{P. Delio is with the Neurology Associates of Santa Barbara and with the Santa Barbara Cottage Hospital (e-mail: prdelio@sbneuro.com).}
\thanks{*A. Parsay and M. Torun contributed equally to this work.}
\thanks{Approval of all ethical and experimental procedures and protocols was granted by the Institutional Review Board (IRB) Committee at UCSB.}
\thanks{This work was supported in part by NSF CNS award 2226255, and in part by ONR award N00014-23-1-2715.}
\thanks{Copyright (c) 2025 IEEE. Personal use of this material is permitted. However, permission to use this material for any other purposes must be obtained from the IEEE by sending a request to pubs-permissions@ieee.org.}
}

\markboth{IEEE INTERNET OF THINGS JOURNAL, VOL.~X, NO.~X, XXXX~XXXX}%
{Parsay \MakeLowercase{\textit{et al.}}: Gait Disorder Assessment Based on a Large-Scale Clinical Trial}

\IEEEpubid{}

\maketitle

\begin{abstract}
Neurological gait disorders affect a large population, significantly reducing life quality. This paper brings a foundational understanding to the potentials of emerging sensing modalities (e.g., WiFi) for gait disorder assessment, via conducting a one-year-long clinical trial in collaboration with the Neurology Associates of Santa Barbara. Our medical campaign encompasses 114 real subjects and a wide spectrum of disorders (e.g., Parkinson’s, Neuropathy, Post Stroke, Dementia, Arthritis). We then develop the first WiFi-based gait disorder sensing system of its kind, distinguished by its scope of validation with a large and diverse patient cohort. To ensure generalizability, we mainly leverage publicly-accessible online videos of gait disorders for training, and develop a video-to-RF pipeline to convert them to synthetic RF training data. We then extensively test the system in a neurology center (i.e., the Neurology Associates of Santa Barbara). Additionally, we provide a 1-1 comparison with a vision-based system, by developing a vision-based gait assessment system under identical conditions, a first-of-its-kind comparison to our knowledge. We finally contrast both systems with neurologists’ accuracy when basing evaluation solely on visual gait inspection, by designing/distributing a large survey to 70 neurologists, offering the first apples-to-apples comparison of these three sensing modalities. Our findings can help integrate these sensing systems into medical practice, working towards equitable healthcare.
\end{abstract}


\section{Introduction} \label{sec: Introduction}
\IEEEPARstart{N}{eurological} gait disorders affect 24\% of adults ages 60 and older, with the number increasing to 55\% above the age of 80 \cite{mahlknecht2013prevalence}. In general, gait disorders can be devastating to the mobility, independence, cognition, and self-esteem of an
individual, as many studies have shown \cite{giladi2013classification}. As such, proper diagnosis is crucial for giving the individual proper care that can optimize their well-being. Yet, many
individuals may not seek medical help, or may seek it only after the disease has progressed \cite{giladi2013classification}. Furthermore, once medical help is sought, many patients miss follow-up appointments \cite{jack2010barriers}. The situation can be much worse in impoverished/developing nations (or even in rural areas within the United States) as families can face additional challenges in accessing healthcare due to prohibitive costs, extended waiting time for specialist consultations, or the general scarcity of medical expertise.

On the other hand, other promising sensing modalities have emerged over the past few years. For instance, wireless signals, now ubiquitous, have created new opportunities for sensing and learning about the environment, causing much enthusiasm
in the area of RF sensing. More specifically, RF signals have been used for applications such as person identification \cite{wang2016gait, zou2018wifishapelet, korany2020multiple, deng2022gaitfi, zheng2019Widar}, crowd analytics~\cite{korany2021counting, cheng2017device, TMC19_DepatlaMostofi}, fall detection~\cite{wang2016wifall,wang2016rt,hu2021defall}, and activity recognition~\cite{chen2018wifi, wang2017device,torun2023wiflex}, among others. Similarly, in the area of vision, the widespread availability of affordable, high-resolution cameras has opened up new sensing possibilities for various applications.   

In this paper, our first major goal is to fundamentally understand the potential of everyday WiFi signals for neurological gait disorder diagnosis.  While RF-based gait disorder analysis has started to gain attraction in the literature, there are still very limited work in this area. For instance, existing WiFi-based ones either do not test with real patients or have only tested with a handful of patients, are limited in the range of considered disorders, and are typically trained and tested in the same environment (i.e., not in the wild) \cite{ zhang2023IoT,tahir2019wifreeze}. On the other hand, there are a small number of papers using radar signals for gait disorder assessment \cite{rana2022markerless,seifert2018radar, yang2022artificial}. However, the majority of these studies are also afflicted by the previous limitations. More importantly, such work use specialized equipment, which can be cost-prohibitive.  In contrast, \textbf{a WiFi-based approach could offer a considerably more economically viable alternative, with the potential to democratize access to healthcare.}  The next section provides a detailed comparison with the prior work, with the key points further summarized in Table \ref{tab:SotA}.
\IEEEpubidadjcol

In the area of vision, on the other hand, there are more work on gait disorder assessment with cameras, due to the prevalence of video datasets that can be used for training/testing purposes. \textbf{Yet, there are no apples-to-apples comparisons between WiFi and video-based systems} -- i.e., evaluations conducted under identical training and testing conditions and with the same subjects. Such a comparison is crucially needed for a proper integration of these systems in medical practices. Our second key objective is then to create a vision-based gait disorder system, mirroring the same subjects/conditions of our WiFi-based one, to facilitate a one-to-one performance comparison.

Finally, a successful integration of such sensing modalities in a smart healthcare system requires rigorous benchmarking against the diagnostic performance of medical experts.  One key role the neurologists play is \textit{the visual assessment of the gait}. Thus, the accurate duplication of this aspect by an RF-based or a vision-based system holds promises for revolutionizing the diagnostic procedure.  In other words, neurologists utilize many inputs (e.g., blood tests, genetic tests, etc) to establish a diagnosis, in addition to a visual inspection of the gait.  However, many of these inputs can be collected more easily (i.e., do not require the specialized expertise of neurologists) and integrated into a smart health system.  Thus, if a smart sensing system can replicate the complex task of visual inspection, it has the potential to automate the diagnostic process, reducing the burden on specialists, and enabling healthcare access for a larger population. 

However, there are no existing work on assessing the diagnostic accuracy of neurologists if only relying on the visual inspection of the gait, motivating our third goal. More specifically, we design a large survey and disseminate it to $70$ board-certified practicing neurologists via partnership with Survey Healthcare Global (SHG) \cite{SHG}, in order to collect their performance when providing a diagnosis only based on the visual inspection of the gait.  This then enables an apples-to-apples comparison with both our WiFi-based and video-based systems.

We next explicitly discuss the contributions of the paper, followed by a detailed state-of-the-art comparison.

\textbf{Statement of Contributions:}

1. Through a partnership with neurologists, we provide a comprehensive analysis of WiFi-based gait disorder assessment by \textbf{running a one-year-long clinical trial encompassing 114 subjects and a wide range of disorders}. These conditions include Parkinson’s disease, Neuropathy, Post Stroke, Dementia, and Arthritis, in addition to healthy subjects. The subjects further exhibit different degrees of severity: mild, moderate, and severe. Finally, the testing is done in the wild, i.e., in the doctor's office. Beyond demonstrating the performance of WiFi signals when detecting gait disorders, the paper further comprehensively shows the impact of several different parameters (e.g., age, height, weight and disease severity) on the performance. It additionally tests with a few samples of other conditions (amyotrophic lateral sclerosis (ALS), multiple sclerosis (MS), spinal cord injury, neurofibromatosis, and dysferlinopathy), as well as cases with multiple disorders. To the best of our knowledge, \textbf{this paper serves as the first in-depth evaluation of WiFi's capability in gait disorder assessment,} with the potential to make a significant leap forward in the field. 

2. To ensure generalizability of our findings and further alleviate the extensive data collection required for training purposes, \textbf{we have trained a neural network for RF-based gait disorder detection by mainly leveraging publicly-accessible online videos of gait disorders/healthy gaits.} This is enabled by our video-to-RF pipeline that converts such videos of a person's walk to the corresponding synthetic RF data, as if the walk was performed near a pair of WiFi transceivers.

3.  Equally important, the paper provides a detailed one-to-one comparison with a vision-based system, by further \textbf{developing a vision-based gait disorder assessment system that uses identical subjects and conditions for both training and testing purposes.}  This side-by-side comparison is crucial towards developing smart health systems and the first of its kind to the best of our knowledge. Furthermore, as will be demonstrated, the findings indicate comparable performance between the two modalities.  

4.  In order to establish the diagnostic accuracy of neurologists when only using visual inspection of the gait, we have designed a large video-based survey (based on the same pool of data) and have further partnered with Survey Healthcare Global (SHG), an entity that provides “measurable healthcare expert opinions,” via recruitment and other tools, who then disseminated our survey to 70 board-certified practicing neurologists.  We then contrast the performance of our WiFi-based and video-based systems with the diagnostic accuracy of neurologists when they base their evaluation solely on the visual inspection of the gait. \textbf{To the best of our knowledge, this is the first result of its kind, which provides an apples-to-apples comparison of these three sensing modalities.} As the results will show, both WiFi and Video-based systems surpass the performance of neurologists when only using visual inspection of the gait. 

Overall, the findings of the paper can provide guidelines for smart health system development, shape the integration of these new sensing modalities into medical practice, and pave the way toward equitable access to health.  \textbf{We note that our IRB committee has reviewed and approved this research.}

\textbf{Remark 1:} Our WiFi CSI dataset, collected from patients with various gait disorders as well as from healthy individuals, is available on our website~\cite{parsay2024wifigait} or upon request to the authors.

\textbf{Remark 2:} For brevity, in the rest of this paper, we refer to the Neurology Associates of Santa Barbara where the experiments took place as \textbf{The Neurology Center}.


\section{Related Work} \label{sec: Related Work}

This section reviews emerging sensing systems for gait-based neurological disorders. It further extensively compares this work with state-of-the-art.  A detailed summary of this comparison is also provided in Table~\ref{tab:SotA} for RF-based systems.  

The prior work can be classified to three main categories:  methods utilizing floor/wearable sensors, video-based techniques, and RF-based approaches.

\textbf{1) Wearable and floor sensors:}
Accelerometers, gyroscopes, and magnetometers have been recently used as floor/wearable sensors for classifying neurological disorders that have a gait manifestation \cite{caldas2017systematic, chen2016toward, tao2012gait, kaur2023deep, pandit2019abnormal, sadiagnosis, banaie2011introduction, zhao2021multimodal}. These sensors are typically attached to the body, e.g., smartwatches and insole pressure sensors, or installed on the ground/treadmill, e.g., force plates. 

Overall, while such recent work can be valuable in better understanding an abnormal gait pattern, incorporating them into practice comes with significant drawbacks, such as limited public accessibility (price and delivery), 
patient discomfort, and difficulty in use \cite{kaur2022vision,mehrizi2019automatic}.

\textbf{2) Video-based Methods:}
With the increasing accessibility and advancements in computer vision techniques, utilizing video footage for gait disorder assessment has become a recent area of exploration \cite{gong2020novel, dentamaro2019real, dentamaro2020gait, verlekar2018automatic, li2018classification, kaur2022vision, mehrizi2019automatic}. For instance, \cite{gong2020novel} develops an R-CNN-based classifier to distinguish between healthy and Parkinsonian gaits, while \cite{dentamaro2020gait} performs classification (health vs. unhealthy), based on 2D pose estimation. Similarly, Verlekar et al. \cite{verlekar2018automatic} use 2D footage of actors, for gait impairment classification. Authors in \cite{li2018classification} classify two gait-affecting neurodegenerative disorders: Parkinson's and Hemiplegia, by utilizing a 3D Kinect sensor.  On the other hand, \cite{jinnovart2020abnormal} classifies healthy vs. unhealthy gaits, considering 5 unhealthy gait disorders that are acted out by 10 subjects. 

In general, the video-based systems, while simpler-to-use as compared to wearables, can raise privacy concerns and/or may not be favorable for some patients that are not comfortable being camcorded. Furthermore, they require a  line-of-sight view. \textbf{Nevertheless, in addition to advancing the state-of-the-art in the area of RF-based gait disorder assessment, this paper further contributes to the area of vision-based gait assessment}, as discussed next.  

In general, there are more papers on vision-based gait disorder assessment, as compared to RF-based approaches.  However, a direct comparison between WiFi-based and vision-based systems, under identical conditions and with the same subjects, is currently lacking. Such a side-by-side methodical comparison is important for providing guidelines on how such systems can be incorporated into the future smart health ecosystem. This paper thus contributes to the area of vision by providing such a comparison.  Furthermore, this paper contrasts the performance of a vision-based system (as well as its WiFi-based counterpart) with that of neurologists when they provide a diagnosis solely based on the visual evaluation of the gait.  

\begin{table*}
  \renewcommand{\arraystretch}{1.25}
  \caption{State-of-the-art comparison of RF-based gait disorder assessment. We note that while we have stated 6 conditions for this paper, we further test with 5 more conditions, but at a smaller scale.}
  \label{tab:SotA}
  
  \begin{tabular}{|c|c|c|c|c|c|c|c|}
    \hline
    \multirow{2}{*}{\small{\textbf{Ref.}}} & \multirow{2}{*}{\small{\textbf{Method}}}  & \small{\textbf{Scope of}}  &  \hspace{-3pt}\small{\textbf{$\#$ of Conditions}}\hspace{-3pt} &  \small{\textbf{Training Data}} &  \small{\textbf{Train/Test in}} &  \multirow{2}{*}{\small{\textbf{Cost}}} &  \multirow{2}{*}{\hspace{-2pt}\small{\textbf{Accuracy}}\hspace{-2pt}}\\
    &   & \small{\textbf{Evaluation}} & \small{\textbf{(incl. Healthy)}}  & \hspace{-2pt}\small{\textbf{Collection Burden}}\hspace{-4pt}& \hspace{-3pt}\small{\textbf{Diff. Environment}}\hspace{-3pt} & &\\ 
 
  \hline
    \footnotesize{\cite{rana2022markerless}} & \footnotesize{Radar (Gait-based)}  & \footnotesize{24 Real Subjects} & \footnotesize{2} & \footnotesize{High} & \footnotesize{No} & \footnotesize{High}& \footnotesize{$94.9\%$} \\
    \hline
    \footnotesize{\cite{seifert2018radar}} & \footnotesize{Radar (Gait-based)}   & \footnotesize{Only acted out} & \footnotesize{5} & \footnotesize{High} & \footnotesize{No} & \footnotesize{High}& \footnotesize{$88.4\%$} \\
    \hline
     \footnotesize{\cite{yang2022artificial}} & \hspace{-1pt}\footnotesize{Radar (Breathing-based)}\hspace{-1pt} &   \footnotesize{53 Real Subjects} & \footnotesize{2} & \footnotesize{Low} & \footnotesize{Yes} & \footnotesize{High}& \footnotesize{84.53\%} \\  
     \hline
    \footnotesize{\cite{guan2019non}} & \hspace{-5pt}\footnotesize{WiFi (Endurance-based)}\hspace{-5pt}   & \footnotesize{Only acted out} & \footnotesize{2} &  \footnotesize{High}& \footnotesize{No} & \footnotesize{Low} & \footnotesize{$99.4\%$} \\    
    \hline
    \footnotesize{\cite{tahir2019wifreeze}} & \footnotesize{WiFi (Gait-based)}   & \footnotesize{15 Real Subjects} & \footnotesize{1} & \footnotesize{High} & \footnotesize{No} & \footnotesize{Low}& \footnotesize{$99.7\%$} \\
    \hline
    \footnotesize{\cite{zhang2023IoT}} & \footnotesize{WiFi (Gait-based)}   & \footnotesize{Only acted out} & \footnotesize{7} & \footnotesize{High} & \footnotesize{Yes} & \footnotesize{Low}& \footnotesize{$94\%$} \\
    \hline    
    \hspace{-2pt}\cellcolor{gray!20}\footnotesize{\textbf{This Paper}}\hspace{-2pt} & \cellcolor{gray!20}\footnotesize{\textbf{WiFi (Gait-based)}}   & \hspace{-2pt}\cellcolor{gray!20}\footnotesize{\textbf{114 Real Subjects}}\hspace{-1pt} & \cellcolor{gray!20}\footnotesize{6} & \cellcolor{gray!20}\footnotesize{Low} & \cellcolor{gray!20}\footnotesize{Yes}  & \cellcolor{gray!20}\footnotesize{Low}& \cellcolor{gray!20}\footnotesize{$\textbf{85.47\%}$} \\ 
  \hline  
\end{tabular}

\end{table*}


\textbf{3) RF-based Methods:}
Radio frequency-based gait disorder analysis has recently gained attraction in the literature.  However, \textbf{there are very limited number of work (six to the best of our knowledge) on gait-disorder assessment using RF signals}, as briefly summarized next. In \cite{rana2022markerless,seifert2018radar}, IR-UWB and K-band radars are utilized to classify healthy vs. abnormal gait, while \cite{yang2022artificial} utilizes a customized frequency-modulated continuous wave (FMCW) radar to detect Parkinson's disease using nocturnal breathing signals. When considering WiFi signals, WiFreeze \cite{tahir2019wifreeze} aims to detect freeze of gait in Parkinson's patients, while \cite{guan2019non} and \cite{zhang2023IoT} classify unhealthy vs. healthy gaits.  Less related to this work is Liu et al. \cite{liu2022monitoring}, which does not detect gait disorders, but analyzes the progress of Parkinson's disease using customized FMCW radio equipment. We next compare our work with existing RF-based gait disorder studies, as also summarized in Table~\ref{tab:SotA}.

\textbf{Scope of evaluation:} Half of the existing RF-based work have not been tested with real patients, rather, the conditions have been enacted \cite{seifert2018radar,guan2019non, zhang2023IoT}. 

For those work that have tested with real patients \cite{rana2022markerless, tahir2019wifreeze, yang2022artificial}, most collect a very small number of patients: 24 (only 4 unhealthy) in \cite{rana2022markerless} and 15 in \cite{tahir2019wifreeze}. \cite{yang2022artificial} has collected data of 53 subjects, albeit with customized radar equipment, and for detection based on breathing signals. In contrast, our paper reports on a comprehensive one-year-long clinical trial involving WiFi-based sensing of 114 subjects, thereby offering a significantly broader scope of investigation.

\textbf{Spectrum of Considered Gait Disorders:} Most existing RF-based gait assessment work have only considered one type of gait disorder (e.g., healthy vs. Parkinson). In this paper, we have undertaken a comprehensive clinical campaign encompassing 5 common gait disorders: Parkinson’s, arthritis, neuropathy, post-stroke, and dementia, plus healthy. In addition, we also test with the following conditions: Amyotrophic lateral sclerosis (ALS), Multiple sclerosis (MS), spinal cord injury, Neurofibromatosis, and Dysferlinopathy, but at a smaller scale.

\textbf{Training Data Collection Burden:} All the existing work (except for \cite{yang2022artificial} which uses breathing signals and customized FMCW radar) have manually collected RF training data, which can be cumbersome and can result in non-generalizable results. This work, however, mainly uses available online videos to generate synthetic RF data for training, thus significantly reducing the training data collection burden and further ensuring generalizable results.

\textbf{Testing in the Wild:} Most existing work (all WiFi-based ones) that collected data of real subjects train and test in the same environment (using the same overall pool), potentially yielding non-generalizable results.  In this paper, on the other hand, we adopt a novel approach by training with synthetic RF data derived from online videos, while testing in The Neurology Center during its normal operating hours. This ensures the robustness and generalizability of the findings of the paper.   
 
Overall, this paper provides a comprehensive analysis of RF-based gait disorder assessment by running a one-year-long clinical trial encompassing a large spectrum of disorders (5 common gait disorders plus healthy).  It then details the results of the data analysis of 114 subjects, while predominantly employing synthetic RF training data generated from available online videos, and relying only on off-the-shelf WiFi devices (a pair of laptops) for sensing.  Equally important, the paper provides a detailed comparison with our developed vision-based system, using identical subjects and conditions for both training and testing. It then contrasts the performance of these systems with the diagnostic accuracy of neurologists when they base their evaluation solely on the visual inspection of the gait. Overall, the findings of the paper can provide guidelines for smart health system development and shape the integration of these new sensing modalities into medical practice, thus paving the way towards equitable healthcare.

\begin{figure}[t]
  \setlength{\abovecaptionskip}{1pt plus 0pt minus 2pt} 
  \centering
  \includegraphics[trim=0mm 3mm 0mm 0mm, clip, width=0.20\textwidth]{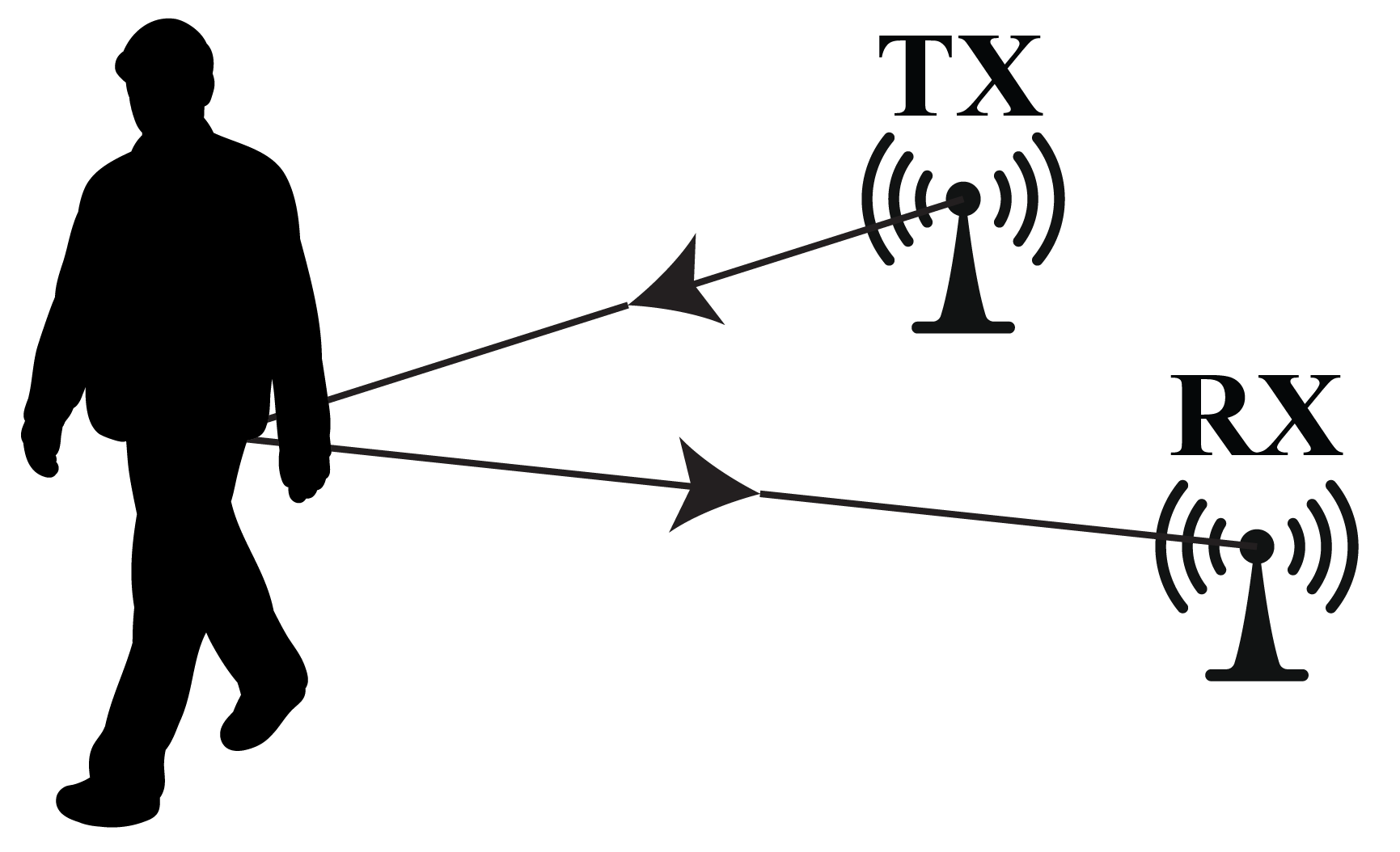}
    \vspace{6pt}
  \caption{The high-level schematic of our WiFi-based gait disorder classification setup.}
  \label{fig: WiFiSystemModel}
\end{figure}

\section{A Primer on RF Signal Modeling} \label{sec: A Primer on RF Signal Modeling}

Consider a person walking in an area, in the vicinity of a pair of WiFi transceivers, as shown in Fig.~\ref{fig: WiFiSystemModel}. The transmitter sends a signal, which bounces off of different parts of the body of the person, as well as other objects in the environment, and is received by the receiver. Then, the complex baseband received signal can be written as:
$s_{\text{rec}}(t) = \alpha_\text{s}e^{j\theta_s}+\sum_{l \in V_{s'}}\alpha_{s',l} e^{j\theta_{s',l}} + \sum_{i \in V_b} \alpha_i e^{j\left(\frac{2\pi}{\lambda}\psi \int v_i(t) dt + \frac{2\pi}{\lambda} d_i\right)}$, where $\lambda$ is the wavelength \cite{korany2019xmodal}. Moreover, $\alpha_i$ is the magnitude of the scattered signal from the $i^{\text{th}}$ body part whose velocity is $v_i(t)$, and $\psi=\cos \phi_{R} + \cos \phi_{T}$, where $\phi_{R}$ and $\phi_{T}$ are the angles between the direction of motion and the vectors from the transmitter and receiver to the body, respectively. $d_i$ is the initial length of the $i^{\text{th}}$ path, and $V_b$ is the set of body points visible to both the receiver and the transmitter. The first and second terms then embrace the impact of the direct path and the reflections from static objects, respectively. We note that at WiFi frequencies, and for the area dimensions considered in this paper, we have a far-field scenario. As such, we took $\psi$ to be the same for all the body parts, as done in the literature. The WiFi receiver, however, can only measure the received signal magnitude or phase difference. We then next summarize an approximated modeling of the received RF signal magnitude \cite{cai2020teaching,korany2019xmodal}. More specifically, since the LOS path is much stronger than the other reflections, i.e., $\lvert \alpha_s\rvert \gg \lvert\alpha_i\rvert$ and $\lvert \alpha_s\rvert \gg \lvert\alpha_{s',l}\rvert$, for $\forall i,l$, we have the following for the received signal magnitude:
\begin{align}
\label{Eq: velocity}
|s_{\text{rec}}(t)| \approx \gamma+\sum_{i \in V_b}\alpha_i \cos(\frac{2\pi}{\lambda}\psi \int v_i(t) dt + \frac{2\pi}{\lambda} d_i-\theta_s).
\end{align}
The first term, $\gamma = \alpha_s+\sum_{l \in V_{s'}}\alpha_{s',l} \cos(\theta_{s',l}-\theta_s)$, is static and removed after DC removal, and the second term embraces the information of the moving body parts. A similar expression can be derived for the phase difference signal.
\begin{figure*}[t!]
  \setlength{\abovecaptionskip}{0pt plus 0pt minus 2pt} 
  \centering
  \includegraphics[width=\linewidth]{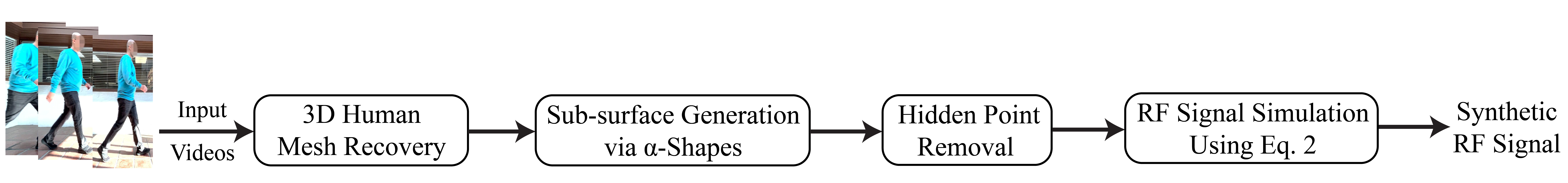}
  \caption{Video-to-RF pipeline to generate synthetic RF training data from public gait disorder videos.}
  \label{fig:systemdesign}
\end{figure*}

\section{Synthetic RF Data Generation} \label{sec: System Design}
\label{sec:systemdesign}
A proper training of a classifier requires extensive training data.  For WiFi sensing applications, however, WiFi sensing datasets pertaining to specific tasks of interest (e.g., gait disorder classification in this paper) are not readily available. Furthermore, manual collection of such datasets can be prohibitive. In the area of vision, on the other hand, there is an abundance of vision dataset pertaining to many different scenarios.  Thus, in recent years, there has been an interest in translating such vision datasets to synthetic RF training datasets (albeit for other applications) \cite{cai2020teaching,torun2023wiflex,xu2022mask}. 

In this paper, we then utilize a publicly-available gait disorder vision dataset of real subjects for generating synthetic RF training data. In this section, we set forth the details of our pipeline for translating a vision input to a synthetic RF signal. As we shall see, the typical human mesh recovery approaches in vision result in a 3D mesh that has a higher sampling rate for certain body parts. This can then magnify the impact of those body parts when generating synthetic RF signals. As such, in this part, we also propose a new approach for remedying this issue. We next discuss the steps in detail.

\subsection{Human Mesh Recovery} \label{sec:HMR}
Consider a video of a person walking with a gait disorder. We first utilize a state-of-the-art human mesh recovery algorithm (HMR) to extract a 3D point cloud of the body for each frame \cite{kanazawa2018end}. However, the output 3D mesh generated by HMR is a non-uniform point cloud whose spatial density is a function of the curvature of the body. Specifically, regions with higher surface curvature, such as the hands, are assigned a denser distribution of mesh points compared to flatter areas like the torso. This distribution is a feature of the SMPL (skinned multi-person linear) model \cite{loper2023smpl}, which serves as the base model in HMR.

\begin{figure}[t]
  \setlength{\abovecaptionskip}{0pt plus 0pt minus 2pt}
  \centering
  \includegraphics[trim=0mm 0mm 0mm 0mm, clip, width=0.35\textwidth]{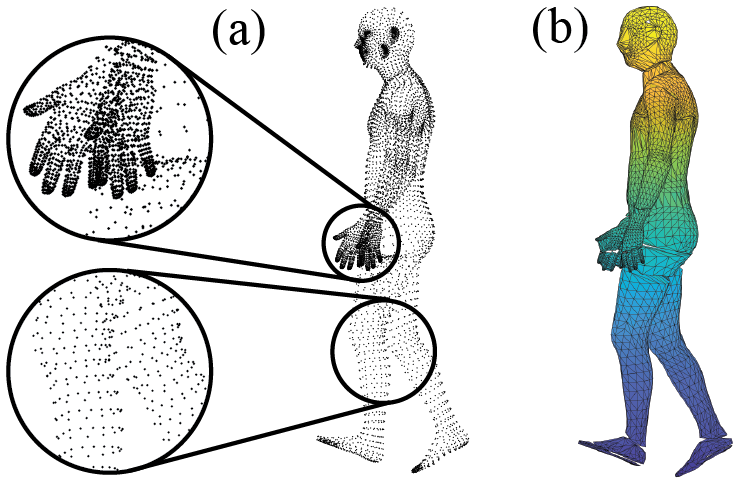}
    \vspace{6pt}
  \caption{Output of (a) HMR and its non-uniform sampling issue and (b) $\alpha$-shape algorithm resolving it using Delaunay triangulation.}
  \label{fig: nonUniformSampling}
  
\end{figure}

Fig.~\ref{fig: nonUniformSampling} (a) shows an example of this where we have utilized state-of-the-art HMR to translate a snapshot of a walk into its 3D point cloud. We can see the zoomed-in mesh points for two body parts: the hands and the legs, where hands are sampled much more heavily due to the high surface curvature. This can then result in the body parts with higher curvatures contributing more than they truly would to the received signal. We next show how to address this issue using $\alpha$-shapes.

\subsection{Finding $\alpha$-Shapes of Human Mesh}\label{sec: alphashape}
HMR provides the 3D mesh as a set of discrete point cloud. We then translate the points to non-overlapping sub-surfaces that cover the whole body using $\alpha$-shapes. The $\alpha$-shape of a set of points is a polytope (not necessarily convex) that best fits the corresponding set \cite{edelsbrunner1994three}.  It can thus be considered as a generalization of convex hulls. By forming the $\alpha$-shape of our given mesh points, we then effectively smooth out the impact of non-uniform sampling and further generate a set of non-overlapping triangular sub-surfaces, for wave simulation purposes.  More specifically, we use the output body partitions provided by HMR (e.g., torso, shoulder, etc.) and create the $\alpha$-shape of each body part, using Delaunay triangulation.  Fig.~\ref{fig: nonUniformSampling} (b) shows the generated $\alpha$-shapes for Fig. \ref{fig: nonUniformSampling} (a). As seen, we now have a set of sub-surfaces whose area is proportional to the area of the corresponding body part. By incorporating the area of each sub-surface into the scattering coefficient as will be discussed in Sec. \ref{sec: ModelingScattering}, we can effectively resolve the original issue caused by the non-uniform sampling rate.

\subsection{Coordinate System Alignment} \label{sec: alignment}

Since HMR provides its output in pixels, the $\alpha$-shapes will also be in pixels. However, for synthetic wave simulation, the positions need to be converted to real-world coordinates. Therefore, by using the provided camera location and walk distance from the video, we then translate the pixels into meters. We further adjust for the angle between the camera and the body centroid at each point along the walking line to address the distortion caused by the viewing angle, similar to \cite{cai2020teaching}. Finally, we place the resulting 3D human shape, consisting of triangular sub-surfaces, in the vicinity of a pair of transceivers in a simulation environment, according to the desired setup, to simulate the walking away motion \cite{cai2020teaching}.


\subsection{Visibility Assessment} \label{sec: visiblity Assessment}

Let $c_i$ and $n_i$ denote the center and normal vector of the $i^{\text{th}}$ sub-surface. We apply a visibility detection algorithm \cite{katz2007direct} to the centers to find sub-surfaces visible to both the transmitter and receiver. With a finite sampling rate, this algorithm may misclassify a small set of points \cite{katz2007direct}. To address this, we perform an additional step. Let $\hat{x}^{(i)}_{inc}$ and $\hat{x}^{(i)}_{RX}$ denote the unit vector from the transmitter to $c_i$, and from $c_i$ to the receiver, respectively. If  $\langle \hat{x}^{(i)}_{inc},n_i \rangle$ or $-\langle\hat{x}^{(i)}_{RX},n_i\rangle$ is positive, then the $i^{\text{th}}$ point will not be visible to the transceivers, where $\langle , \rangle$ denotes the inner product. We then remove all such points if still in the output, thus further reducing the chance of false positives. We denote the final visible set by $V_b$.


\subsection{Modeling the Scattering Pattern} \label{sec: ModelingScattering}

Consider the specular (mirror-like) vector of the $i^\text{th}$ sub-surface: $\hat{x}^{(i)}_{spec}=\hat{x}^{(i)}_{inc}-2\langle\hat{x}_{inc},n_i\rangle n_i$. We model the quasi-specular scattering pattern of the body using a Lambertian model \cite{nayar1991surface, lafortune1997non}, $R(c_i)= \sqrt{A_i \cos(\theta^{(i)}_{inc})}(\alpha e^{j\phi} +\beta cos^m(\theta_i))$, where $\alpha$ and $\beta$ are the relative strength of these terms, $m$ determines the directivity of the quasi-specular component, and $\phi$ is randomly chosen from $[0,2\pi]$. Moreover, $A_i$ is the area of the $i^{\text{th}}$ sub-surface and $\theta_i$ is the angle between $\hat{x}^{(i)}_{RX}$ and $\hat{x}^{(i)}_{spec}$. In this manner, $A_i \cos(\theta^{(i)}_{inc})$ takes the effective area of the $i^{\text{th}}$ surface into account. This further ensures that body surfaces with a higher curvature, which inherently have more sampling points assigned to them (as discussed in Sec.~\ref{sec:HMR}), do not unrealistically affect the total received power.

\subsection{Synthetic Wave Simulation} \label{sec: Synthetic Wave Simulation}
We finally simulate the interaction of the incoming wave with the 3D human shape consisting of the $\alpha$-shape-based sub-surfaces of Sec.~\ref{sec: alphashape}, with the transceivers as discussed in Sec.~\ref{sec: alignment}, and by using the visibility assessment and scattering model of Sec.\ref{sec: visiblity Assessment} and \ref{sec: ModelingScattering}, respectively. More specifically, by using Born approximation, we have  \cite{korany2019xmodal, cai2020teaching}, 
\begin{equation}
    s_{r} = g(P_{TX},P_{RX}) + \sum_{i\in V_b} g(P_{TX},c_i) R(c_i) g(c_i,P_{RX})
    \label{Eq: VideoChannelModel}
\end{equation}
where $g(p_1,p_2)=\frac{1}{4\pi \|p_1-p_2\|}e^{j\frac{2\pi \|p_1-p_2\|}{\lambda}}$ is the free space Green's function, $R(c_i)$ is the scattering coefficient of the $i^{\text{th}}$ sub-surface discussed in Sec. \ref{sec: ModelingScattering}, and $V_b$ is the set of visible points obtained in Sec. \ref{sec: visiblity Assessment}. Eq. \ref{Eq: VideoChannelModel} then allows us to translate a video input to synthetic RF data, which we shall use for training purposes in this paper.

\section{Clinical Trial of WiFi-based Gait-disorder Assessment: Setup}
\label{sec: Experimental Setup and Data Collection}

In this section, we present the details of our one-year-long clinical trial for testing the proposed pipeline. This is then followed by Sec.~\ref{sec:results}, where we extensively show the performance of the proposed pipeline when assessing the gait in The Neurology Center.

\subsection{Experiment Environment} \label{sec:exp_environment}

We have partnered with board-certified neurologists from The Neurology Center, in order to extensively test our proposed pipeline with real patients exhibiting a variety of gait disorders, and in a real office space. Through this partnership, the clinic dedicated a 7m $\times$ 3.18m covered space to our experiments, as shown in Fig.~\ref{fig: experimentSetup}. We have then placed a pair of laptops in this area, where one acts as a WiFi transmitter and the other as a WiFi receiver. 

As discussed in Sec.~\ref{sec: Introduction}, the system is envisioned to be used in smart health settings (e.g., smart home setting).  As such, patients should be asked to walk the most meaningful route for the assessment, which is the one perpendicular to the line connecting the transmitter and receiver. Then, the patients are asked to walk back and forth two times, starting from the marked start point to the marked end point. We note that we do not have any control over how the patients walk, and as such, they can deviate from a straight path at times. We finally note that the route is \textit{not marked} on the ground from the start to the endpoint, as enforcing walking on a marked route can affect how the patients walk, and can thus obscure the gait disorder. We then place the transmitter and receiver 64cm apart (std=16cm), and at the height of 1.1m (std=5cm). The distance from the start point to the link connecting the transmitter and receiver is 95cm (std=15cm). Then, the patients are asked to walk back and forth two times, from the start point to the end point, where each one-way is 3.9m (std=0.1m). We note that the observed slight variability in these parameters is due to the restrictions from The Neurology Center on leaving behind floor markers/equipment.

\textbf{Remark 3:}  In this paper, a gait sample refers to a one-way walk from start to end in Fig.~\ref{fig: experimentSetup}, or vice versa.

\subsection{Duration of the Clinical Trial}

We have run extensive experimental validations with real subjects in the aforementioned Neurology Center over the course of one year. Since some patients may revisit the clinic, we have ensured that each patient is only included once in our experiments.

\subsection{Test Subjects}

In the first phase of this clinical trial, we have tested our proposed pipeline with the WiFi data of $101$ subjects, spanning over 6 gait conditions: Parkinson's disease (20), dementia (14), post-stroke (16), arthritis (hip or knee, 8), neuropathy (10),
and healthy (33). We note that the uneven distribution of the numbers across conditions partly mirrors the visit frequency of each class to the clinic.

We further note that the participants may experience different degrees of disease severity, which can affect the sensing performance.  We extensively discuss such impacts in the next section. Finally, we note that we have also collected additional data from subjects suffering from other conditions, or multiple simultaneous conditions, as detailed in the next section.  

It is important to collect the data of healthy subjects in the same environment for consistency. As such, the healthy individuals are recruited from the following groups: patients who visited the clinic for other neurological issues that do not have a gait manifestation (mainly from migraine patients), the companions of the patients, and the staff of The Neurology Center, among others. 

As mentioned earlier, most patients walked back and forth two times. We then only use those parts where the participant walks away from the link as it is indicated to have a better signal quality \cite{korany2019xmodal}. This then amounts to two gait samples for most patients, each of which then independently contributes to the overall test pool. Overall, 101 subjects amounted to a total of 204 gait samples. 

\begin{figure}[t!]
        \centering
\includegraphics[width=0.8\linewidth,trim=0mm 0mm 0mm 0mm, clip]{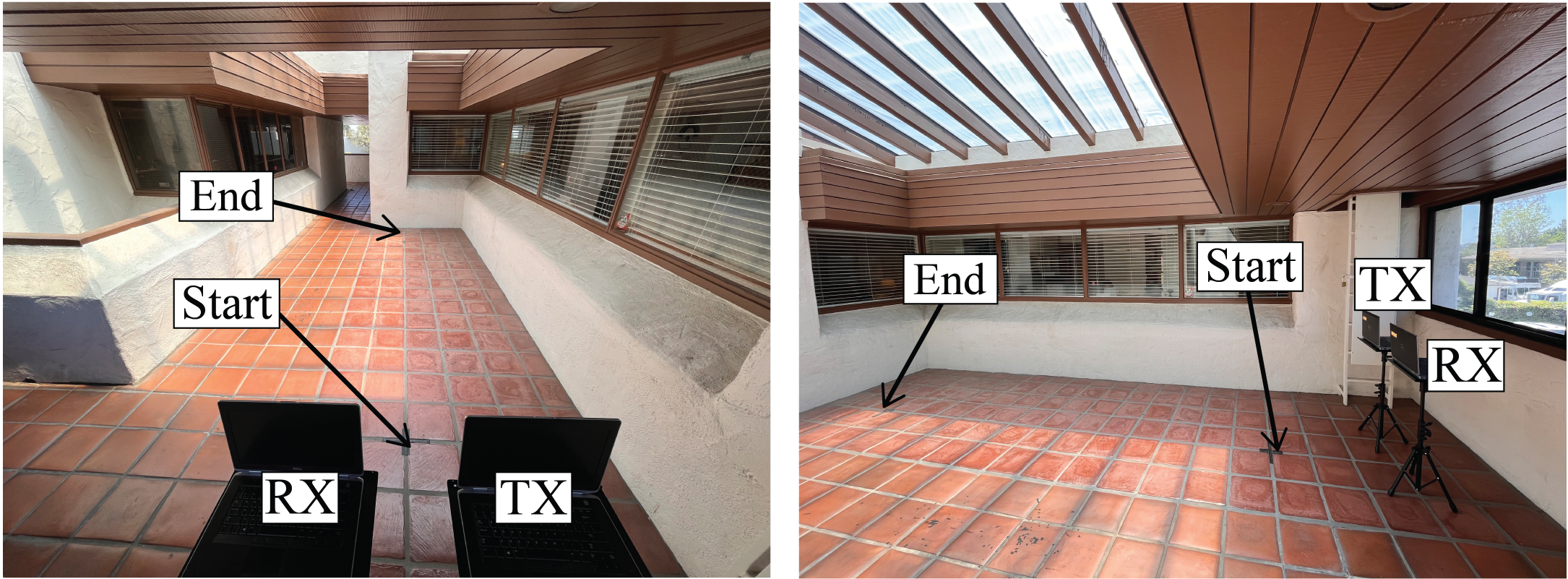}
  \vspace{-4pt}  
  \caption{Two views of our clinical trial area in The Neurology Center.}
  \label{fig:specexamples}
  \label{fig: experimentSetup}
             \vspace{-6pt}
\end{figure}


\subsection{Human-Subject Protocol and Recruitment Process}\label{sec: recruitment}

During their visit to the center, the patients are handed out a pamphlet that details the purpose of the study, the procedure, the compensation, and other details to ensure compliance with the guidelines of our Institutional Review Board (IRB). Once a patient agrees to take part in our medical trials, he/she then walks back and forth a couple of times in the area of Fig.~\ref{fig: experimentSetup}, as detailed in Section~\ref{sec:exp_environment}. The collected data are then stored in an anonymized manner (using an identification code), and the ground-truth gait diagnosis is provided to us by the doctor.  

\textbf{We note that our Institutional Review Board (IRB) committee has reviewed and approved this research.}  

\subsection{Establishing the Ground-truth}\label{sec: groundTruth}

Having a ground-truth for establishing the accuracy of our proposed system is crucial. Neurologists utilize many inputs to establish a diagnosis, in addition to a visual inspection. These can include: the results from vision and speech assessment, sensory perception and mental status evaluations, brain scans, genetic and blood tests, cerebro-spinal fluid analysis, and electroencephalography~\cite{department2019}. Our medical partners then furnish us with their definitive diagnosis for each patient, which we then adopt as the ground-truth.  In addition, the doctor provide us with their assessment of the severity of each condition, categorizing it as mild, moderate, or severe.

\subsection{WiFi Data Collection and Processing}\label{sec: WiFi data collection and processing}

We use a pair of laptops equipped with an Intel 5300 WLAN card as transceivers to collect the WiFi data (See Fig.~\ref{fig: experimentSetup}). While the patient is walking, the transmitter sends 500 packets/second in channel 64 with a center frequency of 5.32 GHz. For each of the three receiving antennas, the channel state information (CSI) of 30 sub-carriers and the subsequent signal magnitude is obtained using CSI Tool \cite{Halperin_csitool}, resulting in 90 data streams. We then denoise the data by applying Principal Component Analysis (PCA). For each PCA component, a spectrogram is generated using a combination of multi-window Hermite functions and STFT, based on a window of size 0.3 seconds and a time shift of 16 ms \cite{korany2019xmodal}. The final spectrogram is then generated by averaging over the first 15 spectrograms.


\subsection{Generating Synthetic RF Training Data from Public Videos}\label{Gen Synthetic Public}

In this paper, we utilize available online videos to generate synthetic RF training data, as proposed in Sec.~\ref{sec:systemdesign}. More specifically, using available online videos \cite{kour2022vision}, we have formed a video dataset of 78 subjects for synthetic WiFi training data generation, encompassing the following three gait conditions: Parkinson's, arthritis, and healthy. This dataset has side-view perspective of the subjects, which is most informative for HMR \cite{kanazawa2018end}. We note that there are no publicly-available dataset for the other three gait disorders of interest to this paper.  However, this has little impact on our performance as we shall see in the next section. Each subject then contributes to two gait samples (except for one person), amounting to a total video dataset of $155$ gait samples: Parkinson's (26), arthritis (70), and healthy (59). 

We then apply the proposed pipeline of Fig.~\ref{fig:systemdesign} to each video. More specifically, the Human Mesh Recovery (HMR) algorithm of~\cite{kanazawa2018end} is first applied to the video to extract a 3D point cloud of the person, which is then up-sampled  to 250 fps. We then apply the $\alpha$-shape algorithm \cite{edelsbrunner1994three} to each body part individually, to generate the sub-surfaces. For each body part, we set a fixed $\alpha$-radius, which we empirically optimize by utilizing a small number of videos. In general, the smaller the body part, the smaller the $\alpha$-shape radius should be.  

After generating the sub-surfaces, we set the WiFi transceivers in a simulation environment, according to the configuration of Fig.~\ref{fig: experimentSetup}, in order to simulate the synthetic RF data that would have been generated if the person in the video was in a WiFi area. We then apply the wave simulator of Sec.~\ref{sec: System Design}, using the reflection model of Sec.~\ref{sec: ModelingScattering} and visibility assessment of Sec.~\ref{sec: visiblity Assessment}. To empirically optimize and set the parameters for $m$, $\alpha$, and $\beta$ of this simulator, we simultaneously collect the WiFi and video data of 2 participants, one with normal and one with an abnormal gait, which are then only used for this purpose (not in training or testing).  Overall, the end-to-end pipeline results in $155$ synthetic RF gait samples of $78$ subjects.  

\subsection{Spectrogram Generation and Feature Extraction} \label{wifi_features}

For each RF training data sample, we generate its spectrogram using a combination of multi-window Hermite functions and STFT \cite{korany2019xmodal}. We note that we only consider the central 50\% time span of each walk, in order to remove the acceleration and deceleration parts as well as any other unrelated movements at the onset or conclusion of the activity. We then extract six features from each spectrogram for training a machine learning pipeline, as discussed next.

In order to find the torso speed, we first sum the spectrogram over frequency for each time instant. We next obtain the shortest frequency window where the sum of the spectrogram values is greater than 50\% of the overall sum (for each time step). This leads to two distinct curves that mark the lower and upper boundaries of the resultant frequency band. We then remove the outliers using the generalized extreme studentized deviate (GESD) test, and smooth the curves by running a moving average over 4 samples. Finally, we apply the same 50\% method to the obtained spectrogram band and average the upper and lower bound curves to get the torso speed. We then find the following six features accordingly: \noindent\textbf{Average torso speed:}  This is the average of the torso speed over time; \noindent\textbf{Minimum and maximum of torso speed:} We calculate the minimum and maximum speed of the torso from its speed curve; \noindent\textbf{Gait Cycle:} This is obtained by finding the peak values (at least 0.3 seconds apart) in the autocorrelation function of the torso speed \cite{wang2016gait, korany2019xmodal}; \noindent\textbf{Step Length:} We calculate the average step length, which is the stride time (half of gait cycle) times the average torso speed \cite{korany2019xmodal}; \noindent\textbf{Torso speed variations:} We obtain the average difference between the maximum torso speed (90\% percentile) and minimum torso speed (10\% percentile) across all steps \cite{korany2019xmodal}.


\subsection{Training a Gait Disorder Classifier for WiFi-based Sensing} \label{RFtraining}

We train a feed-forward, fully connected multi-layer perceptron (MLP) neural network with two hidden layers for WiFi-based gait disorder detection, given the features of Sec.~\ref{wifi_features}, and by using the generated synthetic video-to-RF data of 78 subjects. The hidden layers have 512 and 256 nodes, respectively. The dropout layers are applied to both hidden layers with the parameter 0.5. We utilize ReLU as the activation function for the hidden layers and softmax function at the output layer. Adam optimizer with a learning rate of 0.00001 and a weighted cross-entropy loss function is then used to train the network for 100 epochs, with a batch size of 10.

\subsection{Domain Adaptation}
The synthetic RF training data is generated from videos. Thus, it not only is a completely different modality, but it also naturally encompasses different subjects and environments as compared to those used in testing. These two factors result in the well-known domain and distribution shift problem between the training and test samples in machine learning \cite{koh2021wilds}. To address this, we use WiFi data of 12 subjects (6 healthy, 3 Parkinson's, and 3 arthritis) to re-weigh the loss function via an exhaustive search that aims to maximize the overall system's average accuracy for this small dataset. We then repeat the search process ten times and declare the median of the weight as the final value.  To ensure robustness, the gait assessment's overall performance during testing is then averaged over 8 random sampling of the 12 domain-adaptation subjects, out of the collected WiFi data of 101 subjects, while the rest 89 subjects are left for testing in each round.

We next extensively discuss the performance of the proposed pipeline with real patients.
 

\section{Clinical Trial of WiFi-based Gait-disorder Assessment: Performance Evaluation}
\label{sec:results}

In this section, we present and analyze the outcomes of our one-year-long clinical trial in The Neurology Center, involving 114 subjects. Our analysis is comprehensive as it covers various different gait conditions. We further show the impact of the severity of the disease on the classification accuracy, in addition to demonstrating the impact of age, height, and weight on WiFi-based sensing of gait disorder. Moreover, we show the performance of our system with other gait conditions such as ALS, spinal cord injury, dysferlinopathy, and multiple sclerosis, and demonstrate how they are mainly classified as unhealthy. We finally show the results when testing with patients that suffer from multiple gait conditions.  

Overall, the results shed light on the feasibility of these signals for detecting gait disorders and can provide guidance on how they can be incorporated into future health systems.

\vspace{-4pt}
\subsection{Overall Performance}\label{sec: Overall Performance}

\begin{figure}
    \centering
    \includegraphics[width=0.5\textwidth]{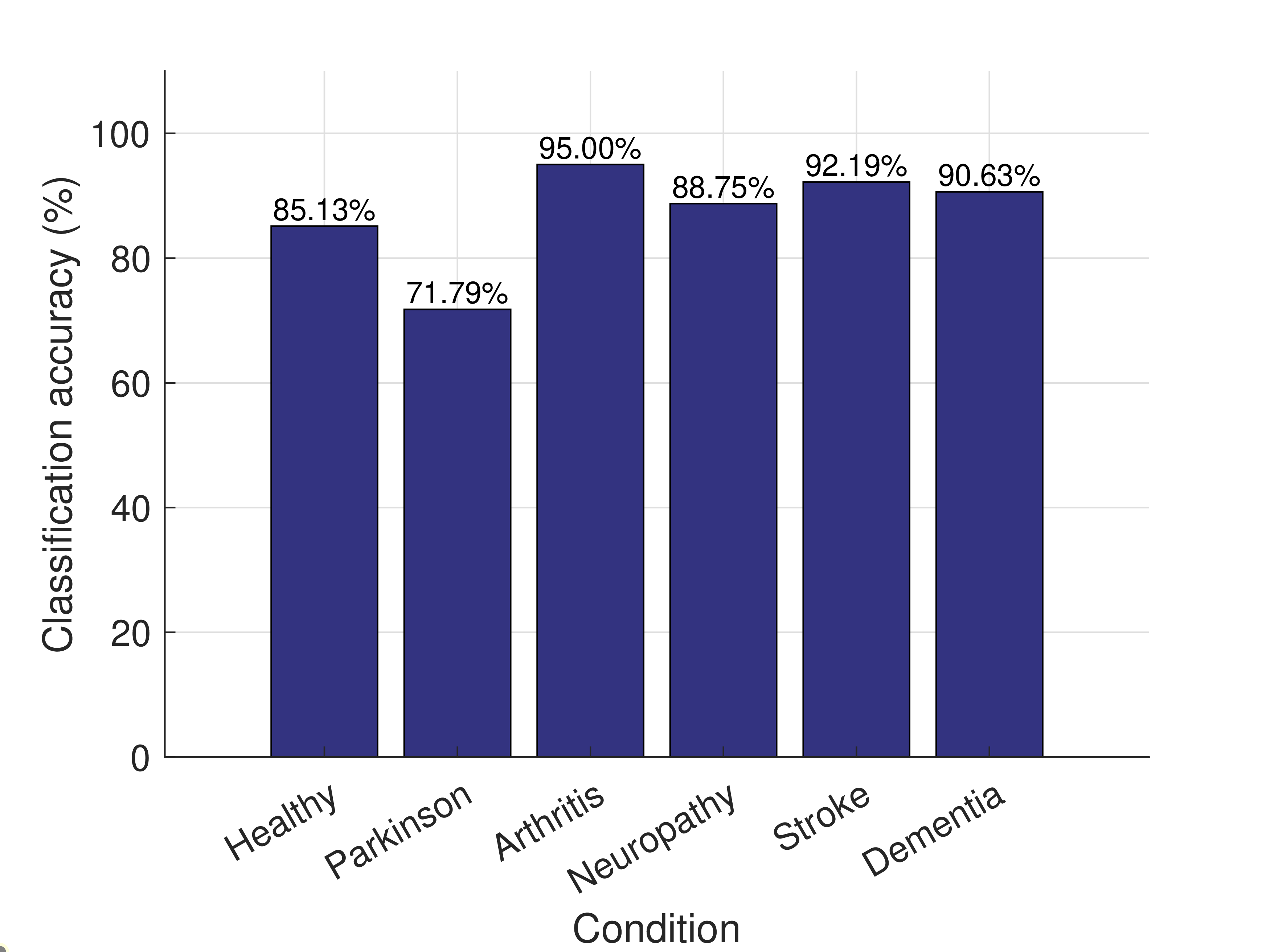}
    \vspace{-14pt}
    \caption{Performance of our WiFi-based system in correctly classifying unhealthy/healthy gaits of 101 subjects in The Neurology Center, with a performance breakdown over each condition. It can be seen that the system can robustly classify unhealthy/healthy gaits when tested extensively with real subjects, with an overall per subject accuracy of 85.61\%, although the system is mainly trained on synthetic data generated from online videos.}
    \label{fig:rf_confmatrix}
    
\end{figure}

In this part, we discuss the performance of our classifier when tested with 101 subjects. More specifically, the overall average per subject accuracy of this classifier is 85.61\%, 
with an average per class accuracy of 85.47\%, with the following breakdown over the two classes: unhealthy (85.8\%) and healthy (85.13\%). Fig. \ref{fig:rf_confmatrix} further shows the average classification accuracy per condition. As can be seen, the performance is mostly consistent per condition. The lower performance for some conditions (e.g., Parkinson’s) can be largely attributed to the disparate distribution of disease severity across conditions, or to how differently these conditions manifest in gait.

\subsection{Impact of Disease Severity}
The severity of a gait disorder can play a key role in its identification, with the expectation that more severe cases have a more clear gait manifestation, making it easier to classify them.  We next show the impact of disease severity on the classification accuracy.  To establish ground truth, The Neurology Center has categorized each case as \textit{mild, moderate, or severe}.  Table \ref{tab: sevAcc} subsequently presents the breakdown of the probability of correct classification (as \textit{unhealthy}) based on disease severity, for the unhealthy gait conditions.   As can be seen, the accuracy increases as the severity of the disease increases, as expected.

\subsection{Impact of Age, Weight, and Height}
\begin{figure*}[t]
  \centering
  \includegraphics[width=0.95\textwidth]{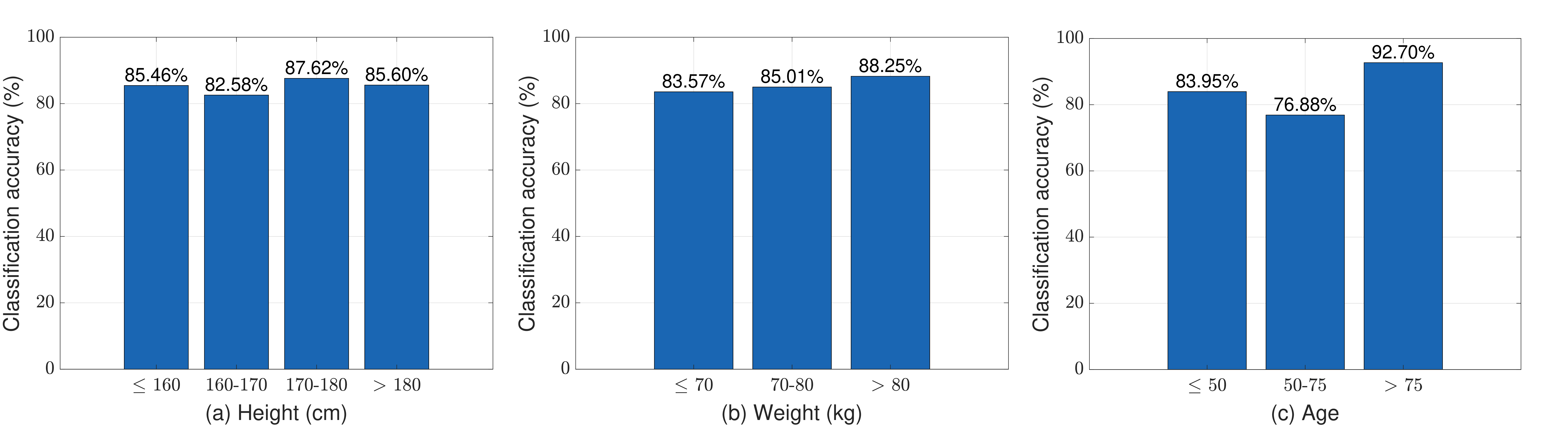}
  \vspace{-4pt}
  \caption{Impact of (a) height, (b) weight, and (c) age on WiFi-based gait disorder assessment.}
  \label{fig:confusionmdemog_vid}

\end{figure*}

This section provides a demographic breakdown of the performance of the proposed classifier, across different age groups, genders, body weights, and heights.  As seen in Fig. \ref{fig:confusionmdemog_vid} (a) and (b), the system shows a consistent performance over different ranges of heights and weights.
Similarly, the classifier attained an accuracy of 86.61\% for individuals self-identifying as males and 84.83\% for those self-identifying as females, demonstrating its capability to accurately classify participants across gender lines. On the other hand, Fig. \ref{fig:confusionmdemog_vid} (c) illustrates enhanced performance in classifying older or younger participants, with a notable decline in the accuracy for middle-aged individuals. We can attribute the observed pattern to the fact that the older subjects typically exhibit more severe symptoms, while the younger group predominantly constitute healthy cases. Meanwhile, the middle-aged cohort generally exhibit symptoms that are less severe than those of older patients, while having a higher incidence of unhealthy cases than the younger group. 

\begin{table}[t]
\caption{Impact of disease severity on correctly classifying it as unhealthy.  The table shows the average overall performance, averaged over all the 5 gait disorder conditions, as a function of disease severity.}

\centering
\renewcommand{\arraystretch}{1.2}
\begin{tabular}{|c|c|c|c|}
\hline
\hspace{7pt}\textbf{Severity}\hspace{7pt} & \hspace{7pt}Mild\hspace{7pt} & \hspace{7pt}Moderate\hspace{7pt} & \hspace{7pt}Severe\hspace{7pt} \\ \hline
\textbf{Accuracy} & 59.38\% & 80.98\% & 100.00\% \\ \hline
\end{tabular}
\label{tab: sevAcc}
\vspace{-4pt}
\end{table}


\subsection{Other Gait Conditions} \label{res:othercond}
We next show the performance of the WiFi-based sensing system when evaluating subjects with other conditions. Towards this goal, we additionally collect WiFi test data for patients with amyotrophic lateral sclerosis (ALS) (1 subject), multiple sclerosis (MS) (3 subjects), spinal cord injury (2 subjects), neurofibromatosis (1 subject), and dysferlinopathy (1 subject) amounting to a total of 8 subjects and $20$ gait samples. It is worth emphasizing that these conditions are not seen during training or domain adaptation. We then feed these to our aforementioned classifier. The system was able to classify all these cases as unhealthy except for two patients, one with ALS (always misclassified) and the other with Neurofibromatosis (one of his/her sample walks was accurately classified in 2 out of the eight random samplings of domain adaptation-test selections while the other was always misclassified).

Overall, this shows the potential of the system to extend to unseen conditions. To improve the accuracy, sample data of these cases can also be included during training and/or domain adaptation.

\subsection{Cases with Multiple Conditions} \label{res:multcond}

Some patients may suffer from multiple gait-related conditions. It is thus important to evaluate the performance of our classifier for such cases. Towards this goal, we have further recruited 5 patients who suffer from multiple conditions: arthritis and neuropathy (1 patient), dementia and post-stroke (1 patient), Parkinson's and neuropathy (1 patient), Parkinson's and Dementia (1 patient), arthritis, neuropathy and spinal cord injury (1 patient) amounting to a total of 10 gait samples.

All these patients were classified as unhealthy by our classifier except for the person with Parkinson's and neuropathy (one sample walk is accurately classified in half of the domain adaption-test selections while the other is accurately classified only in 1 out of 8 selections). Overall, this shows the robust performance of the system even in classifying cases with multiple conditions.


\section{\textbf{Gait Disorder Classification Based on Videos}} \label{sec: VideoFeatures}

In this paper, our goal is to provide a comprehensive investigation of gait disorder sensing with different modalities. Towards this goal, we next delve into developing a video-based gait disorder classifier.  While there are a number of video-based gait disorder classifiers, as surveyed in Sec. \ref{sec: Related Work}, the results of this section are important for the following reason.  In order to effectively compare the performance of WiFi and video-based methods, it is crucial to test them under the same exact conditions (same subjects, environment, etc). In other words, they should be compared based on concurrent video and WiFi recordings of the same subjects. This is important since variability of the subjects can impact the severity of the underlying condition, and variability of the location/time can also result in different sensing qualities, preventing derivations of meaningful conclusions. We next lay out the details of our vision-based system and compare it with the previous WiFi-based pipeline.  

\textbf{Training Dataset:}  We use the same exact publicly-available online video dataset, used for generating synthetic RF data in Sec. \ref{Gen Synthetic Public}, for training the vision-based system.  

\textbf{Test Subjects:} For the purpose of testing, we need to concurrently collect videos of the same subjects that are used for testing the WiFi pipeline. The patient recruitment process is as described in Sec.~\ref{sec: recruitment}, where the patients have the option of participating in only WiFi data collection, or both video and WiFi. For those who agree to both, we then collect their video and WiFi simultaneously, with the video being recorded from the side since it is the most informative for mesh recovery (all online videos are side-views as well). The videos are then recorded using an ultra-wide lens setting, with an aperture of f/2.4 and a 120-degree field of view, at a frame rate of 30 frames per second. We note that the scattered WiFi signals from the tripod and camera are filtered during the DC removal phase, thus not affecting the WiFi pipeline.  \textbf{Finally, we note that our Institutional Review Board (IRB) committee has also reviewed and approved the video collection process.}

\textbf{Feature Extraction and Training:}
The parameters we have used as features in the vision-based system are the same as those utilized in our WiFi pipeline. More specifically, we use the middle 50\% of each walk to exclude the acceleration and deceleration parts. We then generate the 3D point cloud of the body for each frame, which we then partition into 14 parts (hands, legs, torso, etc.). The average position of the points associated with each part results in 14 key points. We obtain the average, minimum (10\% percentile), and maximum (90\% percentile) of the torso speed as our first three features.  To extract the gait cycle, we analyze the vertical position curves of each foot separately and over time. We then filter out the frequencies higher than 4 Hz to remove noise (the cadence for a fast run is around 180 strides/minute, which gives a gait cycle frequency of 1.5 Hz). Subsequently, we determine the gait cycle by computing the time difference between the extrema (at least 0.25 seconds apart) on each curve. Finally, we use the average of two gait cycles as the overall gait cycle, which serves as our fourth feature. It is worth mentioning that if, due to the noise of HMR and lack of visibility of one foot, the algorithm cannot extract the gait cycle from one foot, we only consider the gait cycle of the other foot. Finally, the step length and torso speed variations are obtained similar to our WiFi pipeline.

Using these 6 features, we then train a fully connected neural network with two hidden layers, while keeping all the specifications identical to the WiFi system (see Sec. \ref{RFtraining}). For training, we have the same exact subjects and associated walk samples as in the video-to-RF training dataset of Sec. \ref{Gen Synthetic Public}. We note that while we had 101 subjects for WiFi testing in Sec. \ref{sec: Overall Performance}, we only have consents from 72 of those subjects for video recording. As such, later in this section, we will also further report the performance of the WiFi system with only those subjects as well, for a fair comparison.

More specifically, we have the following breakdown over the 72 subjects: Parkinson’s disease (12), dementia (8), post-stroke (11), arthritis (hip or knee, 4), neuropathy (6), and healthy ($31$), totaling $293$ test gait samples (average of 49 gait samples per condition). Among the patients (10 with mild, 14 with moderate, and 17 with severe gait disorders), 11 subjects (1 with a moderate disorder and 10 with severe disorders) used a walking aid, such as a cane. We note that we use both back and forth of the walks here as they both present informative side-views for a vision-based system. We then use the same number of patients/conditions for domain adaptation (12 total), as described for the WiFi pipeline.  The overall performance is then averaged over 8 random sampling of the 60 subjects out of the overall pool of 72 subjects.  

\begin{figure}[t]
  \centering
  \vspace{-4pt} 
  \includegraphics[width=0.5\textwidth]{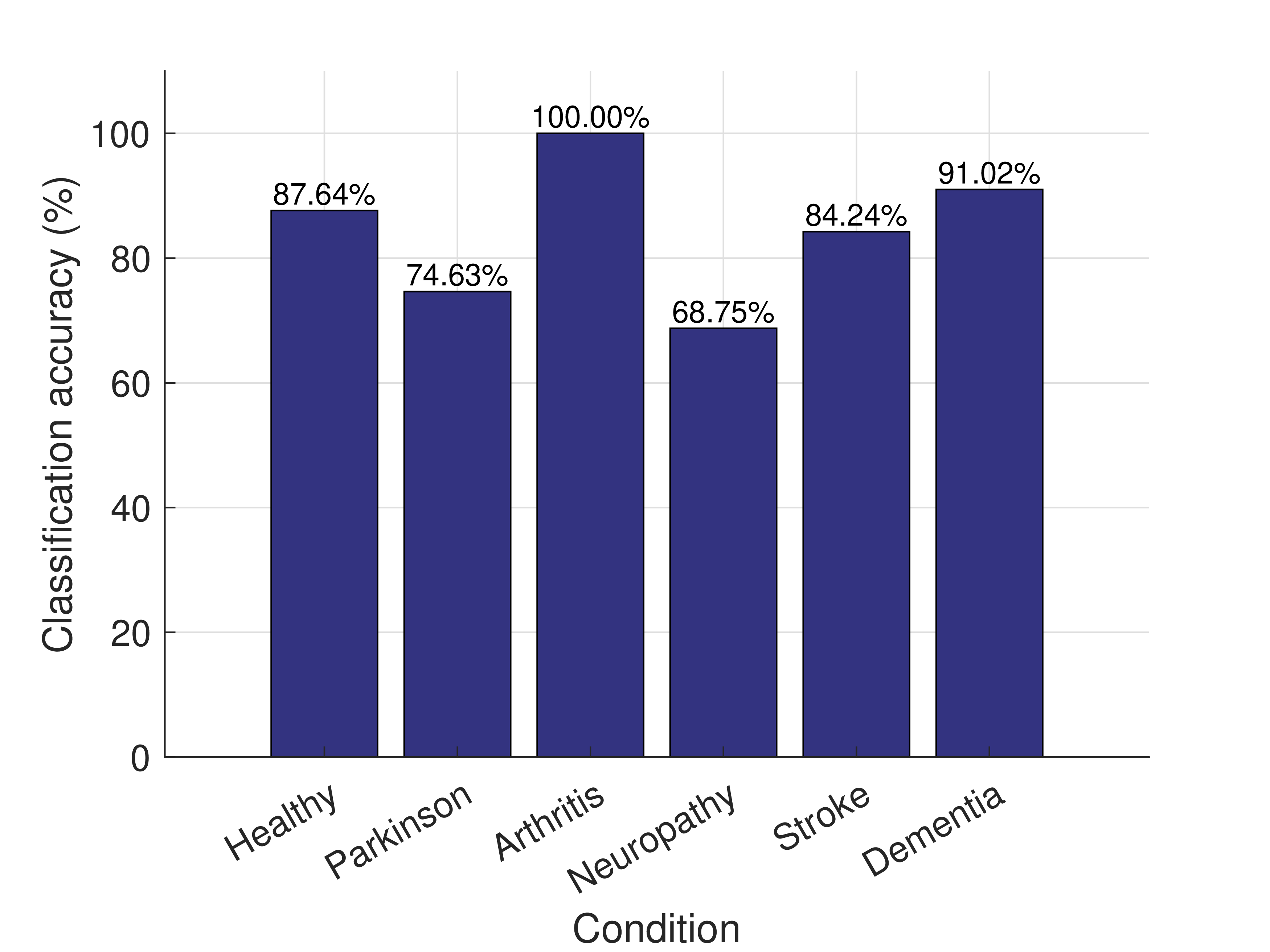}
  \caption{Performance of our video-based system, when tested extensively with real patients exhibiting a spectrum of gait disorders, with an overall average per subject accuracy of $83.80\%$.}
  \label{fig:confusionmTx_vid}
  \vspace{-2pt}
\end{figure}

\textbf{Performance Analysis:} Fig.~\ref{fig:confusionmTx_vid} summarizes the breakdown of the performance over different conditions, for our video-based classifier. Moreover, the overall average per subject accuracy of this classifier is 83.80\%, with an average per class accuracy of 84.37\%, with the following breakdown over the two classes: unhealthy (81.09\%) and healthy (87.64\%). As explained for the WiFi case, the variability of the performance among the conditions can be mainly attributed to the disparate distribution of disease severity across different conditions, or to different gait manifestation of the conditions. Overall, a video-based system can also be a key player in an automated gait disorder assessment system. However, it can raise privacy concerns or may not be favorable for some patients, motivating the use of RF signals. 

\textbf{WiFi-based vs. Video-based:} Fig. \ref{fig:rf_confmatrix} and \ref{fig:confusionmTx_vid} show the results of WiFi and video-based, respectively.  However, as mentioned earlier, we did not have consents of all the 101 subjects for video recording.  As such, while the two systems are trained on the same exact online video subjects, our video pipeline thus far is tested on 72 subjects while the WiFi is tested on a bigger set of 101 subjects.  For a fair comparison, here we report the performance of the WiFi system when only tested on the same exact 72 subjects, while ensuring the same random sampling and domain-adaptation process. This results in an average per class accuracy of 84.41\%, unhealthy (83.03\%) and healthy ({85.78\%), with an overall average per subject accuracy of 84.19\%. We can see that this does not impact the performance of the WiFi system, as compared to the results of Sec. \ref{sec: Overall Performance}, since the number of subjects is high. When comparing this WiFi's performance with that of the video, we can see that they are comparable.  This is promising as it indicates the potential of WiFi for capturing gait disorders.  It {also creates the potential for multi-modality sensing, as we discuss in Sec. \ref{sec: multimodality}. \textbf{To the best of our knowledge this is the first result that compares the performance of WiFi and video when capturing gait disorders, \textit{using the same exact subjects/conditions}}.  Such an apples-to-apples comparison is important for establishing meaningful conclusions.

\section{Medical Experts' Accuracy Based on Visual Inspection }\label{Survey}
As mentioned in Sec.~\ref{sec: Related Work}, neurologists rely on many inputs, in addition to a visual inspection of the gait, to establish a diagnosis. In this paper, we have thus far shown the performance of gait disorder classification based on the information gait induces in the WiFi signal (or in a video). It is thus fitting to find the performance of neurologists when diagnosing a condition based only on the visual inspection of the walk, for proper comparison. \textbf{It is worth noting that this is the first of such a study, to the best of our knowledge.}

\begin{table}[t]
\caption{Neurologists’ accuracy when diagnosing a gait disorder based only on visual inspection.}
\centering

\begin{tabular}{lccc}
\hline
\textbf{Class} & \textbf{Classification Accuracy} \\ \hline
Unhealthy                   & 81.37\%         \\
Healthy                     & 65.95\%          \\
Average per class accuracy                   & 73.66\%\\ \hline
\end{tabular}
\label{tab: survey binary}

\end{table}

Towards this goal, we have designed a survey in Qualtrics \cite{qualtrics}.  Our survey is comprehensive, going beyond binary classification of the gait, which we shall discuss extensively in the next section. In this section, we discuss a part of the survey that facilitates comparison with our developed WiFi and video-based systems. Consider 72 questions (12 per each condition on average). Each question shows a video of a patient's gait and requests a diagnosis by selecting their top three preferred options from a list of six conditions of interest to this paper (the same choice can be made for the top three as well). The videos are from the same subjects used for testing our video-based system for an apples-to-apples comparison. We have further partnered with Survey Healthcare Global (SHG), an entity that provides ``measurable healthcare expert opinions,'' via recruitment and other tools \cite{SHG}, who then disseminated our survey to their pool of certified neurologists. In total, we have received an average of 197 responses per condition from 70 board-certified practicing neurologists, for each of the 6 conditions of interest to this paper (5 gait disorders + healthy). To evaluate the performance of neurologists in diagnosing gait-related conditions, we calculate a binary classification accuracy (healthy vs. unhealthy) similar to the proposed WiFi or video systems. For this purpose, an answer is accurate if a neurologist identifies a healthy (unhealthy) subject as healthy (unhealthy) in their first choice.

Table \ref{tab: survey binary} shows the results for each class (healthy and unhealthy) and the average per-class accuracy. \textbf{It is interesting to observe that both WiFi and video-based systems exhibit slightly superior performance compared to that of neurologists, when they only use visual inspection.}  This is important as it has implications for how WiFi or video-based systems can be incorporated into a smart health system.  As discussed in Sec. \ref{sec: Related Work}, neurologists utilize many inputs when establishing a medical diagnosis. Nevertheless, some of these inputs can be easily collected and integrated with a WiFi or video-based sensing system. Such a system can then provide healthcare for a larger population, and can collaborate with neurologists as needed.   


\section{Discussion and Future Work}
\label{sec:discussion}
We next discuss potential future directions.

\subsection{Can RF Signals Differentiate Neurological Disorders?} \label{sec:probform}
Thus far, we have developed a WiFi-based gait disorder assessment system that can classify the gait into healthy and unhealthy. The natural next step would be to see if such signals are capable of not only identifying an unhealthy gait but also categorizing it to the specific underlying disorder. While there are several work on WiFi signals classifying different activities, such activities typical have a sizable footprint on the received signal.  Exploring the efficacy of these signals in classifying gait disorders thus remains a subject of future exploration. It is worth noting that there are work in the area of vision on classifying the underlying disorder. It would be interesting to see how an RF-based system compares, and further expand the scope of the vision-based systems.  

\begin{table}[t]
\centering
\caption{Neurologists' accuracy when diagnosing solely based on the visual inspection of the gait, with an average per condition accuracy of $30.89\%$.}
\label{tab: surveryAccuracy}

\begin{tabular}{|p{1.4cm}|p{1.5cm}|p{1.5cm}|p{1.5cm}|}
\hline
                      \small{\textbf{Condition}} &    \small{\textbf{Top-1 Acc.}} &  \small{\textbf{Top-2 Acc.}} &   \small{\textbf{Top-3 Acc.}} \\
\hline
   \footnotesize{Arthritis}                                   &                            \footnotesize{18.63\%} &                            \footnotesize{39.83\%} &                            \footnotesize{51.04\%} \\
   \footnotesize{Healthy}                                     &                            \footnotesize{65.95\%} &                            \footnotesize{73.12\%} &                            \footnotesize{80.39\%} \\
   \footnotesize{Dementia}                                    &                             \footnotesize{14.70\%} &                            \footnotesize{30.58\%} &                            \footnotesize{44.29\%} \\
   \footnotesize{Neuropathy}                                  &                            \footnotesize{9.35\%} &                            \footnotesize{23.17\%} &                            \footnotesize{40.78\%} \\
   \footnotesize{Parkinson's}                                 &                            \footnotesize{35.15\%} &                            \footnotesize{43.00\%} &                            \footnotesize{48.11\%} \\
   \footnotesize{Post-stroke}                                      &                            \footnotesize{41.55\%} &                            \footnotesize{51.2\%} &                            \footnotesize{57.19\%} \\
\hline
\end{tabular}

\end{table}

\subsection{Experts' Accuracy Based on Visual Inspection -- Classification} \label{sec: survey discussion}
As outlined earlier, we conducted an extensive survey to evaluate the diagnostic accuracy of neurologists when relying solely on visual gait inspection.  In Sec. \ref{Survey}, we then detailed its findings regarding the classification of gait as either unhealthy or healthy.  Here, we discuss the results when neurologists classify gait into one of the six conditions of interest to this paper. This is important as it provides a benchmark for future RF-based work that aims to further classify the gait disorder to its underlying condition. Table \ref{tab: surveryAccuracy} summarizes the results, with an overall average accuracy of  $30.89\%$ per condition.  These initial results indicate the importance of the other inputs (discussed in Sec. \ref{sec: groundTruth}}) that are part of the diagnostic process, which have the potential to be more easily collected, and integrated with an RF-based or a video-based sensing system. Finally, it is noteworthy that despite using the identical dataset as in Fig. \ref{fig:confusionmTx_vid}, the conditions challenging for assessment differ when assessed by humans.


\subsection{Other Improvements} \label{sec: multimodality}

In this paper, we only used one sample walk instance of the
subject. Using more walk samples or a longer walk duration
can further improve the performance. Moreover, we have
used only 6 features for training the network, for both our
WiFi and video-based systems. Additionally, the same 6 features
are used for both systems for consistency. As part of future work, more features can be utilized and further tailored to each sensing modality. In addition, we have used Born approximation to simulate wave interaction with the body and generate synthetic data from videos.  As part of future work, more advanced wave simulators can be used~\cite{cai2020teaching}, at the cost of an increase in computational complexity. Also, a multi-modal WiFi and video-based system can be developed to increase the robustness. In this paper, we utilized the Linux 802.11n CSI Tool~\cite{Halperin_csitool} for estimating the CSI of different subcarriers, as discussed in Sec.~\ref{sec: WiFi data collection and processing}. Other tools, such as the Atheros CSI Tool~\cite{atherosCSITool} or Nexmon~\cite{Schulz_Nexmon_2017}, or newer WiFi standards can also be employed as part of future work. In addition, using more samples of different gait disorders for training can also improve the quality. For instance, we have only used samples of two gait disorders (plus healthy) during training/domain adaptation. Finally, in terms of disease severity, we only considered \textit{mild, moderate} and \textit{severe} cases.  While considerably challenging, considering \textit{very mild} cases can also be another interesting research avenue.

\subsection{Towards Equitable Healthcare} \label{}

An important future direction would be to integrate additional medical inputs (e.g., patient intake information, blood tests, brain scans) with the collected RF (or video) sensing data.  Given that some of these inputs can be collected more easily (do not require expertise of neurologists), a successful integration can hold promise for an automated smart health system, paving the way towards equitable healthcare.

\section{Conclusions}

In this paper, we reported on the findings of a one-year-long clinical trial for gait disorder assessment. More specifically, we developed the first WiFi-based gait disorder sensing system of its kind, distinguished by its scope of validation with a large and diverse patient cohort. To ensure generalizability, our system was mainly trained on synthetic RF data, converted from publicly-available videos of gait disorders, via developing a video-to-RF pipeline. We further developed a vision-based gait disorder assessment system under the same exact conditions, and provided the first apples-to-apples comparison between the WiFi-based and video-based ones. We finally contrasted both systems with the accuracy of neurologists when basing evaluation solely on the visual inspection of the gait. Overall, the results can play a key role for the proper integration of these sensing modalities into medical practice.


\bibliography{main}
%

\bibliographystyle{IEEEtran}


\newpage

 





\end{document}